\documentclass[10pt]{article}                                                   
\usepackage{sectsty,latexsym,amssymb,dsfont,textcomp}                           
\usepackage[centertags]{amsmath}                                                
\usepackage{fullpage}                                                           
\usepackage[centertags]{amsmath}                                                
\sectionfont{\normalsize}                                                       
\subsectionfont{\normalsize}   
\title{Notes on the Schwinger model: regularization and gauge invariance}
\date{} 

\begin{document}
\baselineskip=10pt
\newcommand{\dirop}{\partial\hspace{-1.25ex}\slash}
\newcommand{\Dirop}{{D}\hspace{-1.55ex}\slash}
\newcommand{\bfU}{\mathbf{U}}
\newcommand{\bfG}{\mathbf{\Gamma}}
\newtheorem{lemma}{Lemma}
\newtheorem{prop}[lemma]{Proposition}
\newtheorem{theorem}[lemma]{Theorem}
\newtheorem{corollary}[lemma]{Corollary}
\newtheorem{definition}[lemma]{Definition}
\newtheorem{remark}[lemma]{Remark}
\newtheorem{Notation}[lemma]{Notation}
\newcommand{\sign}{\hbox{sign}\,}
\newcommand{\eps}{\epsilon}
\newcommand{\proof}{\noindent {\it Proof}\;\;\;}
\newcommand{\qed}{\protect~\protect\hfill $\Box$}
\newcommand{\id}{\mathds{1}}
\newcommand{\be}{\begin{equation}}
\newcommand{\ee}{\end{equation}}
\newcommand{\ba}{\begin{eqnarray}}
\newcommand{\ea}{\end{eqnarray}}
\newcommand{\bes}{\[}
\newcommand{\ees}{\]}
\newcommand{\bas}{\begin{eqnarray*}}
\newcommand{\eas}{\end{eqnarray*}}
\newcommand{\hoa}{{H^{1}_{\mbA}}}
\newcommand{\hta}{{H^{2}_{\mbA}}}
\newcommand{\linf}{{L^\infty}}
\newcommand{\tp}{{\tilde P}}
\newcommand{\cf}{{\cal F}}
\newcommand{\ch}{{\cal H}}
\newcommand{\cfnd}{{\cal F}^{n}-{\cal F}^{n-1}}
\newcommand{\p}{\partial}
\newcommand{\abs}[1]{\vert #1 \vert}

\newcommand{\spsi}{_{{{\Psi}}}}
\newcommand{\pt}{\frac{\partial}{\partial t}}
\newcommand{\pxk}{\frac{\partial}{\partial {x^k}}}
\renewcommand{\theequation}{\arabic{equation}}
\newcommand{\rgt}{\rightarrow}
\newcommand{\lngrgt}{\longrightarrow}
\newcommand{\intsT}{ \int_{0}^{T}\!\!\int_{\Sigma} }
\newcommand{\dxdt}{\;dx\,dt}
\newcommand{\sublt}{_{L^2}}
\newcommand{\sublf}{_{L^4}}
\newcommand{\naf}{\nabla_\mbA \Phi}
\newcommand{\covt}{(\pt -iA_0) }
\newcommand{\ano}{A^{n}_{0}}
\newcommand{\aNo}{A^{N}_{0}}
\newcommand{\Psino}{\Psi^{n}_{0}}
\newcommand{\PsiNo}{\Psi^{N}_{0}}
\newcommand{\Nmo}{{N\!-\!1}}
\newcommand{\nmo}{{n-1}}
\newcommand{\nmt}{{n-2}}
\newcommand{\Nmt}{{N\!-\!2}}
\newcommand{\gotm}{\frac{\gamma}{2\mu}}
\newcommand{\ootm}{\frac{1}{2\mu}}
\newcommand{\tloc}{T_{loc}}
\newcommand{\tmax}{T_{max}}
\font\msym=msbm10
\def\Real{{\mathop{\hbox{\msym \char '122}}}}
\font\smallmsym=msbm7
\def\smr{{\mathop{\hbox{\smallmsym \char '122}}}}
\def\Complex{{\mathop{\hbox{\msym\char'103}}}}
\newcommand{\wkarr}{\; \rightharpoonup \;}
\def\Weak{\,\,\relbar\joinrel\rightharpoonup\,\,}
\newcommand{\To}{\longrightarrow}
\newcommand{\rp}{\hbox{Re\,}}
\newcommand{\pa}{\partial_A}
\newcommand{\pbfa}{\partial_{\mathbf A}}
\newcommand{\pao}{\partial_{A_1}}
\newcommand{\pat}{\partial_{A_2}}
\newcommand{\dbar}{\bar{\partial}}
\newcommand{\barpa}{\bar{\partial}_{A}}
\newcommand{\barpbfa}{\bar{\partial}_{\mathbf A}}
\newcommand{\barpaphi}{\bar{\partial}_{A}\Phi}
\newcommand{\barpbfaphi}{\bar{\partial}_{\mathbf A}\Phi}
\newcommand{\myqed}{\hfill $\Box$}
\newcommand{\cd}{{\cal D}}
\newcommand{\dt}{\hbox{det}\,}
\newcommand{\sma}{_{{ A}}}
\newcommand{\bfpi}{{\mbox{\boldmath$\pi$}}}
\newcommand{\ce}{{\cal E}}
\newcommand{\ulh}{{\underline h}}
\newcommand{\ulg}{{\underline g}}
\newcommand{\ulX}{{\underline {\bf X}}}
\newcommand\la{\label}
\newcommand{\lamo}{\stackrel{\circ}{\lambda}}
\newcommand{\bfjo}{\underline{{\bf J}}}
\newcommand{\Vflato}{V^\flat_0}
\newcommand{\cm}{{\cal M}}
\newcommand{\dist}{{\mbox{dist}}}
\newcommand{\cs}{{\cal S}}
\newcommand{\mcH}{{\mathcal H}}
\newcommand{\mcK}{{\mathcal K}}
\newcommand{\mcF}{{\mathcal F}}
\newcommand{\mcn}{{\mathcal  V}}
\newcommand{\mce}{{\mathcal  E}}
\newcommand{\mcb}{{\mathcal B}}
\newcommand{\mca}{{\mathcal A}}
\newcommand{\mcdpsi}{{\mathcal D}_{\psi}}
\newcommand{\mcd}{{\mathcal D}}
\newcommand{\mcl}{{\mathcal L}}
\newcommand{\mclaphi}{{\mathcal L}_{(\mbA,\Phi)}}
\newcommand{\mcdadjp}{\mcdadj_{\psi}}
\newcommand{\mcdadj}{{\mathcal D}^{\ast}}
\newcommand{\zl}{{Z_\Lambda}}
\newcommand{\thl}{{\Theta_\Lambda}}
\newcommand{\ca}{{\cal A}}
\newcommand{\cb}{{\cal B}}
\newcommand{\cg}{{\cal G}}
\newcommand{\cu}{{\cal U}}
\newcommand{\co}{{\cal O}}
\newcommand{\smA}{\small A}
\newcommand{\ttheta}{\tilde\theta}
\newcommand{\tn}{{\tilde\|}}
\newcommand{\rbar}{\overline{r}}
\newcommand{\oeps}{\overline{\varepsilon}}
\newcommand{\cgl}{\hbox{Lie\,}{\cal G}}
\newcommand{\Ker}{\hbox{Ker\,}}
\newcommand{\AC}{\hbox{AC\,}}
\newcommand{\const}{\hbox{const.\,}}
\newcommand{\Sym}{\hbox{Sym\,}}
\newcommand{\tr}{\hbox{tr\,}}
\newcommand{\grad}{\hbox{grad\,}}
\newcommand{\ttd}{{\tt d}}
\newcommand{\ttdel}{{\tt \delta}}
\newcommand{\ns}{\nabla_*}
\newcommand{\csl}{{\cal SL}}
\newcommand{\kr}{\hbox{Ker}}
\newcommand{\beq}{\begin{equation}}
\newcommand{\eeq}{\end{equation}}
\newcommand{\pr}{\hbox{proj\,}}
\newcommand{\proj}{{\mathbb P}}
\newcommand{\tproj}{\tilde{\mathbb P}}
\newcommand{\projq}{{\mathbb Q}}
\newcommand{\tprojq}{\tilde{\mathbb Q}}
\newcommand{\oN}{\overline N}
\newcommand{\cN}{\cal N}
\newcommand{\cmet}{{{\hbox{${\mathcal Met}$}}}}
\newcommand{\met}{{\hbox{Met}}}
\newcommand{\bfga}{{\mbox{\boldmath$\overline\gamma$}}}
\newcommand{\bfOmega}{\mbox{\boldmath$\Omega$}}
\newcommand{\bfTh}{\mbox{\boldmath$\Theta$}}
\newcommand{\bfmw}{{\bf m}_{\mbox{\boldmath$\smo$}}}
\newcommand{\bfmu}{{\mbox{\boldmath$\mu$}}}
\newcommand{\bfulX}{{\mbox{\boldmath${X}$}}}
\newcommand{\bfultX}{\tilde{\mbox{\boldmath${X}$}}}
\newcommand{\bfmuw}{{\mbox{\boldmath$\mu$}}_{\mbox{\boldmath$\smo$}}}
\newcommand{\dmug}{d\mu_g}
\newcommand{\ltwo}{{L^{2}}}
\newcommand{\lfour}{{L^{4}}}
\newcommand{\hone}{H^{1}}
\newcommand{\honea}{H^{1}_{\smA}}
\newcommand{\er}{e^{-2\rho}}
\newcommand{\onetwo}{\frac{1}{2}}
\newcommand{\lra}{\longrightarrow}
\newcommand{\dv}{\hbox{div\,}}
\newcommand{\mbA}{\mathbf A}
\newcommand{\mba}{\mathbf a}
\newcommand{\mbm}{\mathbf m}
\newcommand{\mbn}{\mathbf n}
\newcommand{\mbx}{\mathbf x}
\newcommand{\opsi}{\overline{\psi}}
\newcommand{\vac}{{\Omega_0}}
\def\smo{{\mbox{\tiny$\omega$}}}
\protect\renewcommand{\theequation}{\thesection.\arabic{equation}}

\font\msym=msbm10
\def\Real{{\mathop{\hbox{\msym \char '122}}}}
\def\R{\Real}
\def\Z{\mathbb Z}
\def\K{\mathbb K}
\def\J{\mathbb J}
\def\L{\mathbb L}
\def\D{\mathbb D}
\def\N{\mathbb N}
\def\Mink{{\mathop{\hbox{\msym \char '115}}}}
\def\Integers{{\mathop{\hbox{\msym \char '132}}}}
\def\Complex{{\mathop{\hbox{\msym\char'103}}}}
\def\C{\Complex}
\font\smallmsym=msbm7
\def\smr{{\mathop{\hbox{\smallmsym \char '122}}}}

\maketitle
\thispagestyle{empty}
\vspace{-0.3in}
\begin{abstract}
The point-splitting computation of the gauge invariant Hamiltonian for
the Schwinger model on the circle in a positive energy representation
is presented.
\end{abstract}

\section{Notation}
\setcounter{equation}{0}
\label{secint}

The starting point is the action functional
\beq\la{act}
S=\int \Bigl[-\frac{1}{4}F_{\mu\nu}F^{\mu\nu}+ 
\overline{\psi}\bigl(i\hbar c\Dirop_A  -\frac{mc^2}{\hbar}  \bigr)
\psi\Bigr]\,dx\,dt
\eeq
describing the interaction of a Dirac spinor field
$\psi$ with an electromagnetic field in 
Minkowski space-time with coordinates $(x^0=ct,\mbx)$ and
metric $c^2dt^2-d\mbx^2$. The electromagnetic field is given 
in terms of the potential $A$ 1-form 
by $F_{\mu\nu}=\partial_\mu A_\nu-\partial_\nu A_\mu$.
The Dirac operator is given by
\beq
\Dirop_A=\gamma^\mu(\partial_\mu-\frac{ie}{\hbar c} A_\mu)
\eeq
where we use the gamma matrices in the form \eqref{gam}.
As in e.g. \cite{sak67} we write
$\opsi=\psi^\dagger\gamma^0$, where $\dagger$ means Hermitian
conjugate.
The equations of motion are
\begin{align}\begin{split}
\partial_\mu F^{\mu\nu}&={-e}\,\overline{\psi}\gamma^\nu \psi\,,\\
i\Dirop_A\psi&=\frac{mc}{\hbar}\psi\,.\end{split}\la{eom}
\end{align}

We work in $1+1$ dimensional
space-time with coordinates $(x^0=ct,x^1=x)$ and
metric $(dx^0)^2-dx^2=c^2dt^2-dx^2$, 
with $0\leq x\leq L$ and periodic boundary conditions. 
Using $x^0$ as the time variable and rescaling
$A,\psi, x$ gives the equations in natural
units (with $\hbar=1=c$). Further, writing $A=ea$ removes the
coupling constant from the interaction so that it appears only
in the denominator in front of the first term in  the action
(and in the commutation relation). As in \cite{mant} we 
will work in the radiation gauge
of Fermi, in which the spatial component of the connection $a$
depends only on time so that the expression for the electric field
$E=\dot a-\partial a_0$ is in fact the decomposition into the
longitudinal and transverse components $E^{long}=-\partial a_0$ and
$E^{tr}=\dot a$ respectively.
The time component $a_0$ is integrated out
via the Gauss law leading to the following classical Hamiltonian 
in the zero mass case:
\beq
H=\int_0^L
\frac{1}{2e^2}{\dot a}^2-\psi^\dagger\bigl(i\gamma^5(\partial-ia)\psi\bigr)
+\frac{1}{2}e^2(\psi^\dagger\psi) (-\Delta)^{-1}*(\psi^\dagger\psi)\, dx\,.
\eeq
Here $(-\Delta)^{-1}$ means the kernel of the operator $-\Delta=-\partial^2$
on $[0,L]$ with periodic boundary conditions, $*$ is convolution and
$\partial=\partial_x$.
Notice that $E^{long}=-\partial a_0$, the longitudinal component of the 
electric field, has been integrated out leaving only the transverse component
$E^{tr}=\dot a$.
We use the following form of the
gamma matrices:
\beq\la{gam}
\gamma^0=\left(\begin{array}{cc}
1 & 0\\
0 & -1
\end{array}\right)\,,\quad
\gamma^1=\left(\begin{array}{cc}
0 & i\\
i & 0
\end{array}\right)\,,\quad
\gamma^5=\gamma^0\gamma^1=\left(\begin{array}{cc}
0 & i\\
-i & 0
\end{array}\right)\,,
\eeq
and we will use dots (resp. $\partial$ or primes) to indicate derivatives with
respect to $t$ (resp. $x$).

In this formulation there is a residual gauge invariance by ``large''
gauge transformations $g_N(x)=e^{2\pi iNx/L}$ for $N\in\Z$, which play an 
important role.

To quantize the theory it is necessary to associate operators
to the fields which satisfy the canonical relations:
\beq\la{car}
\{\psi_\alpha(t,x),\psi_\beta^\dagger(t,y)\}=
\delta_{\alpha\beta}\delta(x-y)
\eeq
(other anti-commutators being zero) and
\beq\la{ccr}
[E^{tr}, a]=[\dot a,a]=-\frac{ie^2}{L}
\eeq
(other commutators being zero). In the process of quantizing 
it is necessary to define carefully what is meant the various
bilinear quantities such as the axial charge
and the Hamiltonian itself.
The end point will be the following formula for the Hamiltonian:
\begin{align}\la{ham}\begin{split}
H=&\;-\frac{e^2}{L}\frac{d^2}{da^2}
+\sum |k_m|(b_m^\dagger b_m+c_m^\dagger c_m)\\
&\qquad a\Bigl(
\sum_{m\geq 0}
{b}^\dagger_{m}b_{m}+
\sum_{m<0}
{b}^\dagger_{m}b_{m}-
\sum_{m> 0} c_m^\dagger c_{m}
-\sum_{m\leq 0}c_m^\dagger c_{m}\Bigr)\\
&\qquad\qquad+\frac{a^2 L}{2\pi}+a
+\frac{e^2L}{2}\sum_{m\neq 0}\frac{1}{k_m^2}\,\jmath_0(-m)\jmath_0(+m)
\,.
\end{split}\end{align} 
Here $b_m^\dagger\,,\, b_m$ (resp. $c_m^\dagger\,,\, c_m$) are fermionic
(resp. anti-fermionic) creation, annihlation operators
acting on the zero charge fermionic Fock space $\mcH_0$, 
with non-interacting vacuum $\Omega_0$,
and $a$ is represented as the operator of co-ordinate multiplication
(Schr\"odinger representation) in the Hilbert space
\beq\la{dhs}
\mcK=\{\Psi=\Psi(a)\in\mcH_0:\Psi\in L^2([0,\frac{2\pi}{L}],da;\mcH_0)
\}\,.
\eeq
Finally the operators $\jmath_0(m)$ are Fourier modes of the current
operator associated to the classical current $j_0=\psi^\dagger\psi$, see
\eqref{ferc0}-\eqref{jm}. The last term in \eqref{ham} is just the
operator associated to the Coulomb energy, 
(which will also require its vacuum expectation to be subtracted.)
 
\section{Solution of the classical Schwinger model}
\setcounter{equation}{0}
The classical equations of motion are
\begin{align}\begin{split}
i\dot\psi&=-i\gamma^5(\partial\psi-ia\psi)-a_0\psi\\
-\Delta a_0
&={-e^2}\,{\psi}^\dagger\psi={-e^2}\,{j^0}\,,\\
-\dot E &={-e^2}\,{\psi}^\dagger\gamma^5\psi={-e^2}\,{j^1}
\,,\end{split}\la{ceom}
\end{align}
where the electric field is $E=F_{01}=\dot a-\partial a_0$. These can be
reduced to linear equations as follows: let $\varphi$ be a solution
of the free Dirac equation
\beq\la{fd}
i\dot\varphi =-i\gamma^5\partial\varphi
\eeq
and write $\psi= e^{i(f+g\gamma^5)}\phi$ with $f,g$ two real-valued
functions, then the first equation of \eqref{ceom} is equivalent
to the pair of equations
\begin{align}\begin{split}
-\dot f&=\partial g-a_0\\
-\dot g &=-a+\partial f
\,.\end{split}\la{ii}
\end{align}
Given $(a_0,a)$ solutions of these may be generated by
first solving the inhomogeneous wave equation
$\ddot f-\partial^2 f=\dot a_0$ and then defining
$g=\int_0^t(a(s,x)+\partial f(s,x))ds$. To complete the reduction
to linear equations we now observe that the currents
${j^0}$ and ${j^1}$ defined in the second two equations
of \eqref{ceom} obey the charge and axial charge conservation laws:
\beq
\partial_\mu j^\mu=0\quad\hbox{and}\quad
\partial_\mu j^{5\mu}=0
\la{cac}
\eeq
where $j=({j^0},{j^1})$ and $j^5=({j^1},{j^0})$.
(These are the conservation laws arising, respectively, from 
phase invariance - under the transformation $\psi\to e^{i\theta}\psi$ ,
and axial phase invariance - under the transformation
$\psi\to e^{i\gamma^5\theta}\psi$.)
\eqref{cac} implies that both ${j^0}$ and ${j^1}$ solve the 
homogeneous wave equation
$$\Box j^\mu=0\,,$$ 
and $a_0,a$ are then determined by solving the second
two equations of \eqref{ceom}. In fact ${j^0},{j^1}$ can be written in 
terms of the free Dirac field $\varphi$ as
${j^0}=\varphi^\dagger\varphi$ and ${j^1}=\varphi^\dagger\gamma^5\varphi$
as a consequence of $(\gamma^5)^\dagger=-\gamma^1\gamma^0=
+\gamma^0\gamma^1=\gamma^5$.
We write $Q=\int {j^0}\,dx$ and $Q^5=\int {j^1}\,dx$ for the corresponding
conserved charges - the first of these is electric charge and the
second will be referred to as axial charge.

\section{\normalsize Quantization of the Dirac field in an external potential}
\setcounter{equation}{0}
Using the infinite Dirac sea representation, second quantization would
associate to the field $\psi$ an operator
$$
\psi=\sum \Bigl(a_n^Ru^Re^{ik_nx}+a_n^Lu^Le^{ik_nx}\Bigr)
$$
where $\{a^{R}_n,a^{R,\dagger}_{n'}\}=\{a^{L}_n,a^{L,\dagger}_{n'}\}=\delta_{nn'}$,
all other anti-commutators being zero; formally this ensures that 
\eqref{car} holds.
The $u^{L,R}$ are eigenvectors of $\gamma^5$ with 
$\gamma^5 u^R=u^R$ and  $\gamma^5 u^L=-u^L$.
Using this representation 
the free Dirac Hamiltonian is 
\beq\la{freek}
\sum k_n(a^{R,\dagger}_{n} a^{R}_n
-a^{L,\dagger}_{n} a^{L}_n)
\eeq 
which fails to be non-negative, or even bounded below.
The physical states are supposed to be those in which all but a finite 
number of negative energy states are occupied. In particular the states
in which all right-moving states with $n\leq P$ and all left-moving states
with $n\geq P+1$ filled are referred to as {\em unexcited} states 
in \cite{mant}. 
In this representation it is necessary to regularize operators such as the 
charge and Hamiltonian in order to get finite expectation values 
on such states;
gauge invariant heat kernel regularization or some variant 
is often used.
Introducing the positive energy representation in which 
fermions and anti-fermions are both explicitly present,
it is possible to avoid using any extreme cut-off such as
heat kernel regularization and instead
use only Schwinger gauge invariant point-splitting to define the bilinear 
quantities. Thus we define

\begin{align}\begin{split}
b_n&=a_n^R\qquad (n\geq 0)\\
b_n&=a_n^L\qquad (n<0)
\end{split}\end{align}
and
\begin{align}\begin{split}
c_n&=a_{-n}^{R,\dagger}\qquad (n> 0)\\
c_n&=a_{-n}^{L,\dagger}\qquad (n\leq 0)
\end{split}\end{align}
so that an unoccupied right (left)  
state of negative (positive)
wave number is now re-interpreted as a state filled by 
a right (left) moving anti-fermion.
The relations \eqref{car}
can be guaranteed by writing
\beq
\psi=\frac{1}{\sqrt{L}}\sum_{n\in\Z} \bigl(b_n u_n e^{ik_nx}+c_n^\dagger v_n e^{-ik_n x}
\bigr)\,,\quad k_n=\frac{2n\pi}{L}
\eeq
with 
\beq\la{carm}
\{b_n,b_{n'}^\dagger\}=
\{c_n,c_{n'}^\dagger\}=
\delta_{nn'}
\eeq
(other anti-commutators being zero) and
\begin{align}\begin{split}
u_n&=u^R\id_{\{n\geq 0\}}+u^L\id_{\{n<0\}}\,,\\
v_n&=u^R\id_{\{n> 0\}}+u^L\id_{\{n\leq 0\}}
\,.\end{split}\la{g5ev}
\end{align}
There is a (non-interacting) vacuum ${\Omega_0}$ and associated states
\beq
\Omega_{\mbm,\mbn}=\Pi\, b_{m_i}^\dagger c_{n_j}^\dagger {\Omega_0}
\eeq
where $\mbm=\{m_i\}_{i=1}^M$ and $\mbn=\{n_j\}_{j=1}^N$
range over finite subsets of $\Z$. Let $\mcF$ be the linear
span of all the $\Omega_{\mbm,\mbn}$, let $\mcF_0\subset\mcF$ be the
zero charge subspace in which there are equal numbers of
fermions and anti-fermions, and finally let
$\mcF_0^P\subset\mcF_0$ be the subspace in which $Q^5=P$ so
that $\mcF_0=\cup_{P\in\Z}\mcF_0^P$, where $Q^5$ is as in \eqref{2qq}.

In this representation the {\em unexcited} states arise 
when $M=N=P\in\Z^+$ as follows. The case $P=0$ corresponds to the 
vacuum ${\Omega_0}$.
For $P\geq 0$
let $m_i=n_i=i$ for $0\leq i\leq P$, and define
the unexcited state
\beq
\Omega_{P+1}\;=\;
\prod\limits_{i=0}^P\;\prod\limits_{j=0}^P\, b_{m_i}^\dagger 
c_{-n_j}^\dagger {\Omega_0}\qquad (P\in\Z^+)\,;
\eeq
for $P<0$ and  $M=N=-P$ let $m_i=n_i=-i$ for $0<i\leq -P$, and define
the unexcited state
\beq
{\Omega_P}\;=\;\prod\limits_{i=-1}^P\;\prod\limits_{j=-1}^P\, b_{m_i}^\dagger 
c_{-n_j}^\dagger {\Omega_0}\qquad 
(P\in\Z^{-}-\{0\})\,.
\eeq

For the free (i.e. in the absence of the potential $a_0,a$) Dirac field,
in this representation,
the charge and axial charge are given by
\begin{align}\begin{split}
Q &=\sum b_n^\dagger b_n-c_n^\dagger c_n\,,\\
Q^5 &=\sum_{n\geq 0} b_n^\dagger b_n-
\sum_{n<0} b_n^\dagger b_n-\sum_{n>0}c_n^\dagger c_n
+\sum_{n\leq 0}c_n^\dagger c_n
\,,\end{split}\la{2qq}
\end{align}
and the Hamiltonian is
\beq
\la{fdh}
H_D^0=\sum |k_n|(b_n^\dagger b_n+c_n^\dagger c_n)
\eeq
of which the $\Omega_{\mbm,\mbn}$ are eigenvectors with eigenvalues
$(\sum_{i=1}^M |k_{m_i}|+\sum_{j=1}^N |k_{n_j}|)\,.$
Thus the $b_n^\dagger$ ($c_n^\dagger$) are interpreted as creation
operators for fermions (anti-fermions) in the state determined
by the wave number $k_n$, with corresponding
annihlation operators  $b_n$ ($c_n$). 

Strictly speaking in obtaining
the expressions above for $H_D^0,Q,Q^5$ 
from the corresponding expressions in terms of the fields
it is necessary to discard an infinite (c-number) sum, which is however
divergent. Having done this,
clearly the above expressions for $H_D^0,Q,Q^5$ define
operators which are well defined on domains containing $\mcF$.
For example, on the unexcited states
$$<P\;|Q|\;P>=0\,,\qquad<P\;|Q^5|\;P>=2P$$
and, considering separately the case $P\geq 0$ and $P<0$ we can verify that
$$<P\;|H_D^0|\;P>=\frac{2\pi}{L}P(P-1)\,,$$
for all $P\in\Z$. (The state $\Omega_1$ contains two particles of zero wave number
and contributes nothing to the energy).

The Hilbert space completions of the various
finite particle subspaces $\mcF,\mcF_0$ and $\mcF_0^P$ 
are written $\mcH=\overline{\mcF}$ and 
\begin{align}\begin{split}
\mcH_0\;&=\overline{\mcF_0}=\,\Ker Q\subset\mcH\,,\\
\mcH_0^P&=\overline{\mcF_0^P}=\Ker Q\cap\Ker (Q^5-2P)\subset\mcH_0\,,
\end{split}
\end{align}
where $Q,Q^5$ are the (unbounded) self-adjoint operators on $\mcH$
determined by the formal expressions in \eqref{2qq} on the domain $\mcF$. 
While $\mcH_0$ and $\{\mcH_0^P\}_{P\in\Z}$
are invariant Hilbert sub-spaces for the evolution $e^{-itH^0_D}$ determined
by the free Dirac Hamiltonian, it will transpire that only $\mcH_0$ is invariant
for the interacting theory even though classically the axial charge
is conserved.

\subsection{Gauge invariance and Schwinger regularization}
As just mentioned it is conventional to discard infinite constants which
arise in the definitions of $Q,H_D^0$ in the second quantization scheme
just outlined.
Nevertheless, in treating the interaction with electromagnetic fields 
it is important to analyze carefully these infinite
sums in order to assure gauge invariance. The first sign that care
is needed comes from considering the action of the large gauge transformations
$g_N$ on the vacuum. The gauge transformation $g_1$ acts classically as
a phase rotation $\psi\to g_1\cdot\psi=\psi e^{2i\pi x/L}$ together with 
$a\to g_1\cdot a=a+2\pi/L$. In the second quantized formalism with an infinte
Dirac sea this induces an shift operator: $a^{L,R}_n\to a^{L,R}_{n-1}$, 
with opposing effects on the kinetic energy \eqref{freek} for the left and right
components. In the positive energy representation a consideration 
of \eqref{carm}-\eqref{g5ev} above and the particle hole
picture leads us to introduce
a modified shift operator acting on the sequence of 
creation and annihlation operators as follows:
\begin{align}\la{gt1}\begin{split}
b_n&\to {{\bfG}} b_n{{\bfG}^{-1}}\,=\,b_{n-1}\,,\; n\neq 0\,,\qquad  b_0\to {{\bfG}} b_0{{\bfG}^{-1}}\,=\,c_1^\dagger
\\
c_n&\to {{\bfG}} c_n{{\bfG}^{-1}}\,=\,c_{n+1}\,,\; n\neq 0\,,\qquad c_0\to {{\bfG}} c_0{{\bfG}^{-1}}\,=\,b_{-1}^\dagger
\end{split}\end{align}
with corresponding relations for the adjoints:
\begin{align}\la{gt2}\begin{split}
b^{\dagger}_n&\to {{\bfG}} b^{\dagger}_n{{\bfG}^{-1}}\,=\,b^{\dagger}_{n-1}\,,\; 
n\neq 0\,,\qquad  b_0^{\dagger}\to {{\bfG}} b^{\dagger}_0{{\bfG}^{-1}}\,=\,c_1
\\
c^{\dagger}_n&\to {{\bfG}} c^{\dagger}_n{{\bfG}^{-1}}\,=\,c^{\dagger}_{n+1}\,,\; 
n\neq 0\,,\qquad c^{\dagger}_0\to {{\bfG}} c^{\dagger}_0{{\bfG}^{-1}}\,=\,b_{-1}
\end{split}\end{align}
To these we append the action on the Dirac (bare) vacuum state:
\beq\la{gt3}
{{\bfG}} \Omega_0\,=\,\Omega_{-1}\,=\,b_{-1}^\dagger c_1^\dagger\Omega_{0}\,.
\eeq
Together these imply 
${{\bfG}}{\Omega_P}=\Omega_{P-1}$ for all $P$. Similarly we obtain a corresponding
modified shift action for ${\bfG}^{-1}$: 
\begin{align}\la{gt1i}\begin{split}
b_n&\to  {{\bfG}^{-1}}b_n{{\bfG}}\,=\,b_{n+1}\,,\; n\neq -1\,,\qquad  b_{-1}\to  
{{\bfG}^{-1}}b_{-1}{{\bfG}}\,=\,c_0^\dagger
\\
c_n&\to  {{\bfG}^{-1}}c_n{{\bfG}}\,=\,c_{n-1}\,,\; n\neq 1\,,\qquad c_1\to  
{{\bfG}^{-1}}c_1{{\bfG}}\,=\,b_{0}^\dagger
\end{split}\end{align}
with corresponding relations for the adjoints:
\begin{align}\la{gt2i}\begin{split}
b^{\dagger}_n&\to  {{\bfG}^{-1}}b^{\dagger}_n{{\bfG}}\,=\,b^{\dagger}_{n+1}\,,\; 
n\neq -1\,,\qquad  b_{-1}^{\dagger}\to  {{\bfG}^{-1}}b^{\dagger}_{-1}{{\bfG}}
\,=\,c_0
\\
c^{\dagger}_n&\to  {{\bfG}^{-1}}c^{\dagger}_n{{\bfG}}\,=\,c^{\dagger}_{n-1}\,,\; 
n\neq 1\,,\qquad c^{\dagger}_{1}\to  {{\bfG}^{-1}}c^{\dagger}_{1}{{\bfG}}\,=\,b_{0}
\end{split}\end{align}
and the relation 
${\bfG}^{-1}\cdot{\Omega_0}=\Omega_1
=b_0^\dagger c_0^\dagger\Omega_0$ 
and
${\bfG}^{-1}\cdot{\Omega_P}=\Omega_{P+1}$ in general.

The gauge transformation acting on the creation annilhlation operators
and the vacuum according to \eqref{gt1}-\eqref{gt3} determines a unitary
transformation on $\mcF_0$ which extends to a unitary transformation 
${\bfG}$ on $\mcH_0$.
This transformation commutes with $Q$ and so preserves $\mcH_0$, but it
does not commute with $Q^5$: for example 
$b_3^\dagger c_2^\dagger b_0^\dagger c_{1}^\dagger{\Omega_0}$ is mapped
into $b_2^\dagger c_3^\dagger c_2^\dagger b_{-1}^\dagger{\Omega_0}$, with the 
chirality reducing by 2. Formally $Q^5{\bfG}^{-1}={\bfG}^{-1}(Q^5-2)$ on $\mcF_0$.

The interpretation
of all these formulae is that large gauge transformations can create and 
annihlate fermion/anti-fermion pairs in a way which seems naively
to change the chiral charge: an anomaly.
We show below that, nevertheless, the Schwinger 
regularization $Q^{5,reg}$ of $Q^5$ is unchanged by large gauge 
transformations.

As a first example of Schwinger regularization consider ${j^1}$: since
$\psi$ and $\psi^\dagger$ are operator valued distributions the
product $\psi^\dagger(x)\gamma^5\psi(x)$ cannot be taken without careful
definition. On the other hand the tensor product $\psi^\dagger(y)\gamma^5\psi(x)$
has an unambiguous meaning as an operator valued distribution on 
the product space, and we can consider the limit $y\to x$ by applying
this to a sequence of test functions supported near the
diagonal $x=y$. In order to assure gauge invariance it is necessary
to insert Schwinger's line integral factor (\cite{sch62}), leading
finally to the tentative definition
\beq\la{td}
j^{1,reg}(x)=\lim_{{\theta}\to 0}\int\,
\psi^\dagger(y)e^{ia(y-x)}\gamma^5\psi(x)\chi_{\theta}(x-y)
\,dy
\eeq
where $\chi$ is a smooth periodic function which is zero outside
of $(-L/4,+L/4)$, has $\int_0^L\chi=1$ and $\chi_{\theta}(x)
={\theta}^{-1}\chi(x/{\theta})$ so that the sequence $\chi_{\theta}$ is an 
approximation to the identity as ${\theta}\to 0$. We now claim that 
the limit on the right hand side of \eqref{td} exists for all such $\chi$
in the following sense: taking the  matrix element between arbitrary 
vectors in $\mcF$ the quantity
\beq\la{j1reg}
\frac{1}{L}\lim_{{\theta}\to 0}\iint\,
\psi^\dagger(y)e^{ia(y-x)}\gamma^5\psi(x)\chi_{\theta}(x-y)e^{-ik_m x}
\,dy\,dx\,
\eeq
is well defined $\forall m\in\Z$ and independent of $\chi$ with the 
properties above. We then define $j^{1,reg}$ by the requirement that
\eqref{j1reg} equals its $m^{th}$ Fourier coefficient
$\jmath_1(m)$ - see \eqref{ferc}.
This seems to be sensible in the sense
that (i) any acceptable definition of $j^{1,reg}$ 
should satisfy this condition, and (ii)
we will obtain an operator which does indeed satisfy this condition.
The same method will then be used to define the Hamiltonian.

Fourier series give the momentum representation of the regularized
axial charge density as
\beq\la{ferc}
j^{1,reg}(x)=\sum{\jmath_1}(m)e^{ik_m x}\,,\qquad
{\jmath_1}(m)=\frac{1}{L}\int j^{1,reg}(x)e^{-ik_m x}\,dx\,.
\eeq
Below we shall compute the matrix elements of the ${\jmath_1}(m)$
as defined by the Schwinger regularization \eqref{td}. The most
important is ${\jmath_1}(0)$, or equivalently the regularized axial charge
$Q^{5,reg}=L{\jmath_1}(0)$, since this contains the anomalous contributions
which are required to restore gauge invariance to the Hamiltonian (which
was broken by the introduction of the vacuum in second quantization).
After presenting these computations we then show
that the other ${\jmath_1}(m)$ are given by their formal expressions
without anomalous contributions.
\subsection{Computation of regularized axial charge}
\la{crq}
We now discuss
the axial charge itself which arises when $m=0$ in \eqref{j1reg}; 
explicitly it is
\begin{align}
\begin{split}
Q^{5,reg}=\frac{1}{L}
\lim_{{\theta}\to 0}\sum_{n,n'}\iint\,
&\bigl({b}^\dagger_{n'} u^\dagger_{n'} e^{-ik_{n'}y}+c_{n'}v_{n'}^\dagger e^{+ik_{n'} y}
\bigr)
e^{ia(y-x)}\gamma^5\,\times\\
&\quad\qquad\bigl(
b_n u_n e^{ik_nx}+c_n^\dagger v_ne^{-ik_n x}
\bigr)
\chi_{\theta}(x-y)
\,dy\,dx\,,
\end{split}\la{dq5}\end{align}
with the understanding that the convergence is in the weak operator sense. 
In particular consider the matrix elements between two vectors 
$\Omega_{\tilde\mbm,\tilde\mbn}$ and $\Omega_{\hat\mbm,\hat\mbn}$ in $\mcF$.
Recall that vectors $\Omega_{\mbm,\mbn}\,\,\in\mcF$ are 
annihlated by all but a finite
number of the $b_n, c_{n'}$: this ensures that
expressions with annihlation operators on the right and
creation operators on the left reduce to finite sums. 
In fact the only collection of terms 
inside $\sum_{n,n'}$ which do not collapse to a finite
sum for this reason are those involving $c_n$ and $c_{n}^\dagger$, as all others
are identically zero. For example the terms
arising from the $c_{n'}$ and $b_{n}$ are
$$
\lim_{{\theta}\to 0}\sum_{n,n'}\iint\,
<\Omega_{\tilde\mbm,\tilde\mbn}\,|c_{n'}b_n |\,\Omega_{\hat\mbm,\hat\mbn}>
v_{n'}^\dagger\gamma^5u_n
e^{ia(y-x)} e^{+ik_{n'} y}
 e^{ik_nx}
\chi_{\theta}(x-y)
\,dy\,dx
$$ 
which is zero because $v_{n'}^\dagger\gamma^5u_n=0$
unless $n,n'$ are either both positive or both negative in which
case, writing $z=y-x$ and changing variables,
we have
$$
\iint\,
e^{ia(y-x)} e^{+ik_{n'} y}
 e^{ik_nx}
\chi_{\theta}(x-y)
\,dy\,dx
=
\iint\,
e^{ia(z)} e^{+ik_{n'} (z)}
\chi_{\theta}(z)
 e^{i(k_n+k_{n'})x}
\, dz\,dx=0\,.
$$
Using $\{c_{n'},c_n^\dagger\}=\delta_{nn'}$ the collection of terms
involving $c_n$ and $c_{n}^\dagger$
can be rearranged to have annihlation operators on the right at the 
expense of a c-number term $C_A$. To be precise we end up with
\begin{align}
Q^{5,reg}&=
\sum_{n\geq 0}
{b}^\dagger_{n}b_{n}-
\sum_{n<0}
{b}^\dagger_{n}b_{n}-
\sum_{n> 0} c_n^\dagger c_{n}
+\sum_{n\leq 0}c_n^\dagger c_{n}
+C_A\,,\\
&\hbox{where}\qquad C_A=\frac{1}{L}
\lim_{{\theta}\to 0}\sum_{n}\iint\,
v_{n}^\dagger\gamma^5v_n
 e^{+ik_{n} y}e^{ia(y-x)}e^{-ik_n x}
\chi_{\theta}(x-y)
\,dy\,dx\,.
\notag\end{align}
Introducing the periodic distributions defined by
\begin{align}\begin{split}
\delta_+(z)&=\frac{{1}}{L}\sum_{n>0} e^{ik_n z}
=\frac{{1}}{L}\lim_{\eps\downarrow 0}\frac{1}{e^{\frac{2\pi}{L}(\eps-iz)}-1}
=\frac{{1}}{L}\lim_{\eps\downarrow 0}\frac{1}{e^{\frac{-2i\pi}{L}(z+i\eps)}-1}\\
\delta_-(z)&=\frac{{1}}{L}\sum_{n<0} e^{ik_n z}
=\frac{{1}}{L}\lim_{\eps\downarrow 0}\frac{1}{e^{\frac{2\pi}{L}(\eps+iz)}-1}
=\frac{{1}}{L}\lim_{\eps\downarrow 0}\frac{1}{e^{\frac{2i\pi}{L}(z-i\eps)}-1}\;,
\end{split}\end{align}
so that $\delta(z)=\frac{1}{L}+\delta_+(z)+\delta_-(z)$ is the
$L$-periodic $\delta$ function, we can write
$$
C_A=\lim_{{\theta}\to 0}
\iint\,
\bigl(\delta_+(y-x)-\delta_-(y-x)-\frac{1}{L}\bigr)
e^{ia(y-x)}
\chi_{\theta}(x-y)
\,dy\,dx\,.
$$
The first $\delta_{\pm}$ terms reduce to 
$$\lim_{\theta\downarrow 0}\lim_{\eps\downarrow 0}
\int\frac{1}{e^{\frac{2\pi}{L}(\eps\mp iz)}-1}e^{iaz}\chi_\theta(z)\,dz=
\lim_{\theta\downarrow 0}\lim_{\eps\downarrow 0}(I^\pm_{\eps,\theta}+
II^\pm_{\eps,\theta})$$ where
\begin{align}\begin{split}
I^\pm_{\eps,\theta}&=\int\frac{e^{iaz}-1}{e^{\frac{2\pi}{L}(\eps\mp iz)}-1}\chi_\theta(z)\,dz\\
II^\pm_{\eps,\theta}&=\int\frac{1}{e^{\frac{2\pi}{L}(\eps\mp iz)}-1}\chi_\theta(z)\,dz\,.
\end{split}\end{align}
For even $\chi$ it is clear from parity considerations that 
$II^+_{\eps,\theta}-II^-_{\eps,\theta}=0$. For $I^\pm_{\eps,\theta}$ notice that
$\frac{e^{iaz}-1}{e^{\frac{2\pi}{L}(\eps\mp iz)}-1}\to 
\frac{e^{iaz}-1}{e^{\frac{\mp 2\pi iz}{L}}-1}$ as $\eps\downarrow 0$ 
for all non-zero $z$ in the support of $\chi_\theta$. But also
$$
\left|\frac{e^{iaz}-1}{e^{\frac{2\pi}{L}(\eps\mp iz)}-1}\right|\leq
\frac{|e^{iaz}-1|}{e^{2\pi\eps/L}|\sin\frac{2\pi z}{L}|}
\leq\frac{c|z|}{|\sin\frac{2\pi z}{L}|}\leq c'
$$
in the support of $\chi_\theta$, so using first the bounded convergence theorem
and then the approximation to the identity theorem (\cite[theorem 8.15]{foll})
we end up with:
$$
\lim_{\theta\downarrow 0}\lim_{\eps\downarrow 0}I^\pm_{\eps,\theta}
=\lim_{\theta\downarrow 0}\int\frac{e^{iaz}-1}
{e^{\mp \frac{2\pi iz}{L}}-1}\chi_\theta(z)dz
=\mp\frac{aL}{2\pi}
$$
leading finally to the following formula for the regularized
axial charge operator:
\beq\la{q5r}
Q^{5,reg}=
\sum_{n\geq 0}
{b}^\dagger_{n}b_{n}-
\sum_{n<0}
{b}^\dagger_{n}b_{n}-
\sum_{n> 0} c_n^\dagger c_{n}
+\sum_{n\leq 0}c_n^\dagger c_{n}
-\frac{aL}{\pi}-1\,.
\eeq
On the unexcited states we have
\begin{align}
<P\;|Q^{5,reg}|\;P>&=2P-\frac{aL}{\pi}-1\,.
\end{align}
Notice the dependence on the background connection $a$,
the independence of $a$ of the naive expression for $Q^5$ notwithstanding.

\subsection{Computation of regularized kinetic energy}

We consider the Schwinger regularization of the kinetic energy term
$H^0_D=\int\psi^\dagger(-i\gamma^5\partial)\psi$, taking as starting
point the definition analogous to \eqref{dq5}:
\begin{align}
\begin{split}
H_D^{0,reg}=\frac{1}{L}
\lim_{{\theta}\to 0}\sum_{n,n'}\iint\,
&\bigl({b}^\dagger_{n'} u^\dagger_{n'} e^{-ik_{n'}y}+c_{n'}v_{n'}^\dagger e^{+ik_{n'} y}
\bigr)
e^{ia(y-x)}\gamma^5\,\times\\
&\quad\qquad\bigl(
b_n u_n k_n e^{ik_nx}-c_n^\dagger v_nk_n e^{-ik_n x}
\bigr)
\chi_{\theta}(x-y)
\,dy\,dx\,,
\end{split}\la{dk}
\end{align}
with the limit to be understood in the same sense as in \eqref{j1reg}.
The structure is the same as \eqref{dq5} and by the same reasoning
we have
\beq
H_D^{0,reg}=
\sum_{n\geq 0}
k_n{b}^\dagger_{n}b_{n}-
\sum_{n<0}
k_n{b}^\dagger_{n}b_{n}+
\sum_{n> 0} k_nc_n^\dagger c_{n}
-\sum_{n\leq 0}k_nc_n^\dagger c_{n}
+C'_A
\eeq
Concentrating on the anomalous c-number term, we are led to
consider
\begin{align}\notag\begin{split}
C'_A & =\lim_{{\theta}\to 0}
\iint\,
\bigl(\delta'_+(y-x)-\delta'_-(y-x)\bigr)
ie^{ia(y-x)}
\chi_{\theta}(y-x)
\,dy\,dx\\
&=\lim_{{\theta}\to 0}
\iint\,
\bigl(\delta_+(y-x)-\delta_-(y-x)\bigr)
\bigl(ae^{ia(y-x)}
\chi_{\theta}(y-x)
-ie^{ia(y-x)}\chi_\theta'(y-x)\bigr)
\,dy\,dx\,.
\end{split}\end{align}
The contribution arising from the first term in the second
bracket is $-a^2L/\pi$ by the previous calculation. For
the contribution from the second term: first, we can 
subtract off an $a$ independent c-number and replace 
$e^{iaz}$ by $e^{iaz}-1$, and then calculate the result to be
\begin{align}\begin{split}
&\lim_{{\theta}\to 0}\lim_{\eps\to 0}
-i\int\,
\bigl(\frac{2ie^{\frac{2\pi\eps}{L}}\sin\frac{2\pi z}{L}}
{e^{\frac{4\pi\eps}{L}}+1-2e^{\frac{2\pi\eps}{L}}\cos\frac{2\pi z}{L}}\bigr)
(e^{iaz}-1)\chi_\theta'(z)
\,dz\\
&=\lim_{{\theta}\to 0}\lim_{\eps\to 0}
-i\int\,
\bigl(\frac{2ie^{\frac{2\pi\eps}{L}}\sin\frac{2\pi z}{L}}
{e^{\frac{4\pi\eps}{L}}+1-2e^{\frac{2\pi\eps}{L}}\cos\frac{2\pi z}{L}}\bigr)
\bigl(e^{iaz}-1-iaz\bigr)\chi_\theta'(z)
\,dz\\
&=\lim_{{\theta}\to 0}\lim_{\eps\to 0}
\int\,
\bigl(\frac{\sin\frac{2\pi z}{L}}
{1-\cos\frac{2\pi z}{L}}\bigr)
\bigl(e^{iaz}-1-iaz\bigr)\chi_\theta'(z)\,dz\\
&=-\int\,
\frac{d}{dz}\biggl[\bigl(\frac{\sin\frac{2\pi z}{L}}
{1-\cos\frac{2\pi z}{L}}\bigr)
\bigl(e^{iaz}-1-iaz\bigr)\biggr]\chi_\theta(z)\,dz
\end{split}
\end{align}
using, respectively, parity, bounded convergence theorem and
integration by parts. The final limit can now be evaluated
using again \cite[theorem 8.15]{foll} to be $a^2L/(2\pi)$.
Adding this to the first term we end up with $C_A'=-a^2L/(2\pi)$
and so
\begin{align}
\la{krf}\begin{split}
H_D^{0,reg}
&=
\sum_{n\geq 0}
k_n{b}^\dagger_{n}b_{n}-
\sum_{n<0}
k_n{b}^\dagger_{n}b_{n}+
\sum_{n> 0} k_nc_n^\dagger c_{n}
-\sum_{n\leq 0}k_nc_n^\dagger c_{n}
-\frac{a^2L}{2\pi}\,,\\
&=\sum_{n\in\Z}\,|k_n|\,({b}^\dagger_{n}b_{n}+
c_n^\dagger c_{n})\,-\,\frac{a^2L}{2\pi}\,.
\end{split}
\end{align}
The total Dirac Hamiltonian is
$H_D^a=H_D^{0}-aQ^5$ so all together we posit the following formula
for the regularized Hamiltonian:
\begin{align}
H_D^{a,reg}&=\,H_D^{0,reg}-aQ^{5,reg}\\
&=\,\sum_{n\geq 0}
(k_n-a){b}^\dagger_{n}b_{n}-
\sum_{n<0}
(k_n-a){b}^\dagger_{n}b_{n}+
\sum_{n> 0} (k_n+a)c_n^\dagger c_{n}
-\sum_{n\leq 0}(k_n+a)c_n^\dagger c_{n}
+\frac{a^2L}{2\pi}+a\,,\notag\\
&=\sum_{n\in\Z}\,|k_n|\,({b}^\dagger_{n}b_{n}+
c_n^\dagger c_{n})\,-\,\frac{a^2L}{2\pi}-\,a Q^{5,reg}\,.
\la{krf2}
\end{align}
On unexcited states:
\begin{align}
<P\;|H_D^{a,reg}|\;P>&=\frac{2\pi}{L}P(P-1)
-\frac{a^2L}{2\pi}-a\Bigl(
2P-\frac{aL}{\pi}-1\Bigr)
=\frac{2\pi}{L}\Bigl(
P-\frac{aL}{2\pi}-\frac{1}{2}
\Bigr)^2-\frac{\pi}{2L}\notag\\
&=
\frac{\pi}{2L}\Bigl[\bigl(<P\;|Q^{5,reg}|\;P>\bigr)^2-1\Bigr]
\,.
\end{align}

\subsection{Action of large gauge transformations}

Under the gauge transformation $g_1$ the regularized axial
charge transforms to
$$
\sum_{n\geq 0}
{b}^\dagger_{n}b_{n}
+c_1c^\dagger_1-
\sum_{n<-1}
{b}^\dagger_{n}b_{n}-
\sum_{n> 1} c_n^\dagger c_{n}
+\sum_{n\leq 0}c_n^\dagger c_{n}
+b_{-1}b_{-1}^\dagger
-\frac{(a+\frac{2\pi}{L})L}{\pi}-1\,.
$$
which equals $Q^{5,reg}$ by the anti-commutation relations.

Similarly the regularized kinetic energy transforms to
\begin{align}\notag
&\sum_{n\geq 1}
k_n{b}^\dagger_{n-1}b_{n-1}-
\sum_{n<0}
k_n{b}^\dagger_{n-1}b_{n-1}+
\sum_{n> 0} k_nc_{n+1}^\dagger c_{n+1}
-\sum_{n< 0}k_nc_{n+1}^\dagger c_{n+1}
-\frac{(a+\frac{2\pi}{L})^2L}{2\pi}\\\notag
&\quad =
\sum_{n\geq 0}
(k_n+\frac{2\pi}{L}){b}^\dagger_{n}b_{n}-
\sum_{n<-1}
(k_n+\frac{2\pi}{L}){b}^\dagger_{n}b_{n}+
\sum_{n> 1} (k_n-\frac{2\pi}{L})c_{n}^\dagger c_{n}
-\sum_{n\leq 0}(k_n-\frac{2\pi}{L})c_{n}^\dagger c_{n}
-\frac{(a+\frac{2\pi}{L})^2L}{2\pi}\\\notag
&\qquad = 
H_D^{0,reg}+k_{-1}b_{-1}^\dagger b_{-1}-k_1c_1^\dagger c_1
+\frac{2\pi}{L}\Bigl(
\sum_{n\geq 0}
{b}^\dagger_{n}b_{n}-
\sum_{n<-1}
{b}^\dagger_{n}b_{n}-
\sum_{n> 1} c_{n}^\dagger c_{n}
+\sum_{n\leq 0}c_{n}^\dagger c_{n}
\Bigr)-2a-\frac{2\pi}{L}\\
&\qquad\quad = 
H_D^{0,reg}
+\frac{2\pi}{L}\Bigl(
\sum_{n\geq 0}
{b}^\dagger_{n}b_{n}-
\sum_{n\leq -1}
{b}^\dagger_{n}b_{n}-
\sum_{n\geq 1} c_{n}^\dagger c_{n}
+\sum_{n\leq 0}c_{n}^\dagger c_{n}
\Bigr)-2a-\frac{2\pi}{L}
\end{align}
Since $aQ^{5,reg}$ transforms to $(a+\frac{2\pi}{L})Q^{5,reg}$
it follows that $H_D^{a,reg}$ transforms to
$
H_D^{0,reg}-aQ^{5,reg}=H_D^{a,reg}\,,
$
as required.

\subsection{Computation of ${\jmath_1}(m)$}
We return to the regularized current $j^{1,reg}(x)=\sum{\jmath_1}(m)e^{ik_m x}$
from \eqref{ferc}. We have already computed 
\beq\la{jm0}
\jmath_1(0)=\frac{1}{L}Q^{5,reg}=\frac{1}{L}\Bigl[
\sum_{n\geq 0}
{b}^\dagger_{n}b_{n}-
\sum_{n<0}
{b}^\dagger_{n}b_{n}-
\sum_{n> 0} c_n^\dagger c_{n}
+\sum_{n\leq 0}c_n^\dagger c_{n}
-\frac{aL}{\pi}-1\Bigr]
\eeq
in section \ref{crq}.
The corresponding formulae for the other fourier modes are:
\begin{align}\la{jmm}
\jmath_1(m)=
&\frac{1}{L}\Bigl[\,\bigl(\sum\limits_{n\geq 0 \;\hbox{and}\; n'\geq 0}-
\sum\limits_{n<0\;\hbox{and}\; n'<0}\bigr)\,\delta_{n'+m-n,0}\,b_{n'}^\dagger b_n \\
&\qquad\qquad\bigl(\sum\limits_{n> 0 \;\hbox{and}\; n'\geq 0}-
\sum\limits_{n\leq 0\;\hbox{and}\; n'<0}\bigr)\,\delta_{n'+m+n,0}
\,b_{n'}^\dagger c_n^\dagger \notag\\
&\qquad\qquad\qquad\bigl(\sum\limits_{n\geq 0 \;\hbox{and}\; n'> 0}-
\sum\limits_{n<0\;\hbox{and}\; n'\leq 0}\bigr)\,\delta_{n'-m+n,0}\,c_{n'} b_n \notag\\
&\qquad\qquad\qquad\qquad\bigl(\;\sum\limits_{n\leq 0 \;\hbox{and}\; n'\leq 0}-
\sum\limits_{n>0\;\hbox{and}\; n'>0}\bigr)\,\delta_{n'-m-n,0}\, c_{n}^\dagger c_{n'}
\,\Bigr]\,.
\notag\end{align}
As already mentioned, there are no anomalous contributions to 
the ${\jmath_1}(m)$, as defined by the Schwinger regularization, for
$m\neq 0$. This
is because the $\{c_{n'},c_{n}\}=\delta_{n'n}$ anti-commutator gives
rise to a term involving the integrals
$
\int\, e^{-ik_mx}\int\,
v_{n}^\dagger\gamma^5v_n
 e^{+ik_{n} y}e^{ia(y-x)}e^{-ik_n x}
\chi_{\theta}(x-y)
\,dy\,dx\,
$
which are all identically zero for $m\neq 0$.
Thus it remains to analyze the contributions from the terms
involving the creation annihlation operators, and as before
considering the matrix elements between two vectors 
$\Omega_{\tilde\mbm,\tilde\mbn}$ and $\Omega_{\hat\mbm,\hat\mbn}$ in $\mcF$,
so that the vectors $\Omega_{\mbm,\mbn}$ are annihlated by all but a finite
number of the $b_n, c_{n'}$.

We introduce similarly the regularized charge density by the
formula analogous to \eqref{td} with momentum representation:
\beq\la{ferc0}
{j^{0,reg}}(x)=\sum{\jmath_0}(m)e^{ik_m x}\,,\qquad
{\jmath_0}(m)=\frac{1}{L}\int {j^{0,reg}}(x)e^{-ik_m x}\,dx\,.
\eeq
Calculating as in section \ref{crq} leads to
\beq\la{j0}
\jmath_0(0)=\frac{1}{L}Q^{reg}=\frac{1}{L}\Bigl[
\sum_{n\geq 0}
{b}^\dagger_{n}b_{n}+
\sum_{n<0}
{b}^\dagger_{n}b_{n}-
\sum_{n> 0} c_n^\dagger c_{n}
-\sum_{n\leq 0}c_n^\dagger c_{n}\Bigr]\,,
\eeq
(i.e. there is no anomaly in ${j^0}$) and
the corresponding formulae for the other fourier modes are:
\begin{align}\la{jm}
\jmath_0(m)=
&\frac{1}{L}\Bigl[\,\bigl(\sum\limits_{n\geq 0 \;\hbox{and}\; n'\geq 0}+
\sum\limits_{n<0\;\hbox{and}\; n'<0}\bigr)\,\delta_{n'+m-n,0}\,b_{n'}^\dagger b_n \\
&\qquad\qquad\bigl(\sum\limits_{n> 0 \;\hbox{and}\; n'\geq 0}+
\sum\limits_{n\leq 0\;\hbox{and}\; n'<0}\bigr)\,\delta_{n'+m+n,0}
\,b_{n'}^\dagger c_n^\dagger \notag\\
&\qquad\qquad\qquad\bigl(\sum\limits_{n\geq 0 \;\hbox{and}\; n'> 0}+
\sum\limits_{n<0\;\hbox{and}\; n'\leq 0}\bigr)\,\delta_{n'-m+n,0}\,c_{n'} b_n \notag\\
&\qquad\qquad\qquad\qquad\bigl(\;-\,\sum\limits_{n\leq 0 \;\hbox{and}\; n'\leq 0}-
\sum\limits_{n>0\;\hbox{and}\; n'>0}\bigr)\,\delta_{n'-m-n,0}\, c_{n}^\dagger c_{n'}
\,\Bigr]\,.
\notag\end{align}

Notice that all these expressions give rise to finite operator sums
when matrix elements between elements of $\mcF$ are computed - in fact
the middle two lines define finite sums while the first and fourth,
although in general unbounded, map the finite particle subspace to itself.

\end{document}